\documentclass{emulateapj}

\shortauthors{CHIBA ET AL.}
\shorttitle{Mid-IR Imaging of Quadruple Lenses}

\begin{document}

\title{Subaru Mid-infrared Imaging of the Quadruple Lenses
PG1115$+$080 and B1422$+$231: Limits on Substructure Lensing
\altaffilmark{1}}

\author{Masashi~Chiba\altaffilmark{2},
        Takeo~Minezaki\altaffilmark{3},
        Nobunari~Kashikawa\altaffilmark{4},
        Hirokazu~Kataza\altaffilmark{5},
    and Kaiki~Taro~Inoue\altaffilmark{6}}

\altaffiltext{1}{Based on data collected at Subaru Telescope,
which is operated by the National Astronomical Observatory of Japan.}
\altaffiltext{2}{Astronomical Institute, Tohoku University,
Aoba-ku, Sendai 980-8578, Japan; chiba@astr.tohoku.ac.jp}
\altaffiltext{3}{Institute of Astronomy, School of Science, University of
Tokyo, Mitaka, Tokyo 181-0015,
Japan; minezaki@mtk.ioa.s.u-tokyo.ac.jp}
\altaffiltext{4}{National Astronomical Observatory, Mitaka, Tokyo 181-8588,
Japan; kashik@optik.mtk.nao.ac.jp}
\altaffiltext{5}{Institute of Space and Astronautical Science,
Japan Aerospace Exploration Agency, Sagamihara, Kanagawa
229-8510, Japan; kataza@ir.isas.jaxa.jp}
\altaffiltext{6}{School of Science and Engineering, Kinki University,
Higashi Osaka 577-8502, Japan; kinoue@phys.kindai.ac.jp}

\begin{abstract}
We present mid-infrared imaging at 11.7 $\mu$m for the quadruple lens systems,
PG1115$+$080 and B1422$+$231, using the cooled mid-infrared camera and
spectrometer (COMICS) attached on the Subaru telescope.
These lensed QSOs are characterized by their anomalous optical and radio flux
ratios as obtained for \rm{(A1, A2)} images of PG1115$+$080 and
\rm{(A, B, C)} images of B1422$+$231, respectively, i.e., such flux ratios are
hardly reproduced by lens models with smooth mass distribution. Our mid-infrared
observations for these images have revealed that the mid-infrared flux ratio
A2$/$A1 of PG1115$+$080 is virtually consistent with smooth lens models (but
inconsistent with the optical flux ratio), whereas for B1422$+$231,
the mid-infrared flux ratios among \rm{(A, B, C)} are in good agreement with
the radio flux ratios. We also identify a clear infrared bump
in the spectral energy distributions of these QSOs, thereby indicating that
the observed mid-infrared fluxes originate from a hot dust torus around a
QSO nucleus. Based on the size estimate of the dust torus, we place
limits on the mass of a substructure in these lens systems, causing
the anomalous optical or radio flux ratios. For PG1115$+$080, the mass of
a substructure inside an Einstein radius, $M_E$, is $\la 16 M_\odot$,
corresponding to either a star or a low-mass CDM subhalo having the mass of
$M_{100}^{\rm SIS} \la 2.2 \times 10^4 M_\odot$ inside radius of 100~pc
if modeled as a singular isothermal sphere (SIS). For B1422$+$231,
we obtain $M_E \ga 209 M_\odot$,
indicating that a CDM subhalo is more likely, having the mass of
$M_{100}^{\rm SIS} \ga 7.4 \times 10^4 M_\odot$.
\end{abstract}


\section{INTRODUCTION}

The cold dark matter (CDM) scenario for structure formation in the Universe has
been quite successful to explain a wide variety of observational results, such
as the characteristic fluctuations of the cosmic microwave background (CMB) and
large-scale structure of galaxy distribution, on the spatial scales larger than
$\sim 1$ Mpc. In particular, the recently released WMAP results indicate that
the power spectrum of the CMB radiation matches the CDM prediction remarkably
well (Spergel et al. 2003). The currently best world model, which accords with
the results of many cosmological observations, consists of approximately
23 \% CDM, 4 \% baryon, and 73 \% vacuum energy.

However, the recent advent of high-resolution N-body simulations on CDM-based
structure formation has enabled to highlight various discrepancies with existing
observations on the spatial scales smaller than $\sim 1$ Mpc. One of the most
serious issues is that CDM models predict the existence of several hundred dark
satellites or ``CDM subhalos'' (with masses of $M \sim 10^{7-9}$ M$_\odot$) in
a galaxy-sized halo (with $M \sim10^{12}$ M$_\odot$), in sharp contrast to the
observed number of about a dozen Milky Way satellites (e.g., Klypin et al. 1999;
Moore et al. 1999). Although this discrepancy may be alleviated by some baryonic
processes, such as the suppression effect of background UV radiation on the
formation of visible dwarf galaxies (e.g., Bullock, Kravtsov, \& Weinberg 2000),
the presence of many dark subhalos in a galaxy is inevitable if CDM models are
correct. It is worth noting that the abundance of CDM subhalos is closely
relevant to the small-scale power spectrum of initial density fluctuations,
which determines the formation of the first (Population III) stars and resultant
reionization history of the Universe (e.g., Yoshida et al. 2003). Also, the
abundance of such subhalos may be related to the formation of bright
galaxies like the Milky Way, as their clustering history seems to affect
the final morphological type of forming galaxies
(e.g., Kauffmann, White, \& Guiderdoni 1993).
Thus, clarifying the presence or absence of CDM
subhalos in a galaxy like the Milky Way is closely related to how visible parts
of the Universe have formed.

Metcalf \& Madau (2001) and Chiba (2002) demonstrated that gravitational
lensing is the most powerful probe of such numerous CDM subhalos which reside
in external lensing galaxies. They showed that if these subhalos exist within
lensing galaxies responsible for multiple imaged QSOs, the magnification
properties of lensed images are modified compared to the predictions of smooth
lens models -- as was first noticed by Mao \& Schneider (1998) who examined
the effects of non-smooth density distribution in lenses.
In particular, the anomalous image flux ratios observed in some four-image QSOs,
which are hardly reproduced by any lens models with smooth density distribution,
can be explained by the existence of numerous CDM subhalos in the lensing
galaxies, while the image positions are little affected (Chiba 2002).
More extensive, statistical analyses of anomalous image flux ratios have provided
important limits on the nature of lens substructures and also the power spectrum
of CDM models at small spatial scales 
(e.g., Metcalf \& Zhao 2002; Dalal \& Kochanek 2002; Keeton 2003;
Kochanek \& Dalal 2004; Oguri 2004).
However, such a lensing approach to probing numerous CDM subhalos is subject
to generic uncertainties associated with the microlensing by stellar populations
in lensing galaxies, which causes the same magnification effect on the QSO images
(Schechter \& Wambsganss 2002). Thus, to map out the CDM subhalos alone,
we require more convincing observational studies in addition to the existing
image data.

One of the best techniques to distinguish microlensing by stars
from lensing by CDM subhalos\footnote{This lensing is often referred to as
{\it millilensing} because its Einstein radius is generally of the order of
milliarcsec.} is to
observe flux ratios from emission regions of different size, $R_S$,
and to compare them with 
the mass-dependent size of the Einstein radius,
which is, for a point-mass lens model, given as
$R_E (M) \sim 0.01 (M/M_\odot)^{1/2} h^{-1/2}$~pc (where
$h=H_0/100$ km~s$^{-1}$~Mpc$^{-1}$) at the typical redshift of a QSO
at $z_S = 2-3$ and a lens at $z_L = 0.3-0.5$. If $R_S \gg R_E(M)$ for
either a star or subhalo with mass $M$, then the image is
unaffected by such a lens perturber, whereas it is significantly magnified
if $R_S < R_E(M)$. This indicates that the observed optical continuum emitted
from a QSO nucleus, a central engine of an accretion disk
with an emission region of $R_S \sim 10^{15}$~cm
(e.g., Wambsganss, Schneider, \& Paczy\'nski 1990; Rauch \& Blandford 1991;
Wyithe et al. 2000), is subject to lensing magnification by both stars and
CDM subhalos. On the other hand, if we focus on
spatially extended regions around the central engine,
such as a surrounding dust torus responsible for infrared emission of
a QSO (e.g., Sanders et al. 1989) or line-emitting regions (e.g., Moustakas
\& Metcalf 2003), then stars would be unable to affect
the flux from such extended regions of $R_S \ga 1$ pc
(Agol, Jones, \& Blaes 2000; Metcalf et al. 2004).
Thus, this selective magnification offers an important probe for
distinguishing the effects of CDM subhalos from those of stars on the concerned
QSO images showing anomalous optical flux ratios.

In this paper, we report the mid-infrared observations of lensed QSOs showing
anomalous optical flux ratios using the Subaru telescope.
The observed mid-infrared waveband
corresponds to the near-infrared waveband in the rest frame,
and its flux is considered to be dominated by thermal radiation
from hot dust located at the innermost region of a dust torus
(Barvainis 1987; Kobayashi et al. 1993).
The inner radius of a dust torus would be determined
by the highest sublimation temperature of dust ($T\sim 1800~K$)
and the UV luminosity of a QSO central engine,
and it is estimated as $R_S\sim 1$~pc for
UV luminosity of $10^{46}$ erg~s$^{-1}$ (Barvainis 1987).
Therefore, the observed mid-infrared flux is
totally insensitive to microlensing by foreground stars ($R_S \gg R_E$)
but is strongly magnified by CDM subhalos ($R_S \ll R_E$).

Moreover, mid-infrared observations of lensed images are advantageous
compared to other wavebands in this kind of study:
(1) mid-infrared flux is almost free from the effects of
differential extinction among different images, in contrast to the use
of continuum or line fluxes in shorter wavebands,
(2) imaging data of both radio-quiet and radio-loud QSOs are available,
and
(3) the source size of mid-infrared flux,
which originates from near-infrared emission in the rest frame,
is quantitatively estimated from the rest-frame
optical luminosity of a QSO
based on the result of the dust reverberation observation, i.e.
the measurement of arrival time between UV flux from a central engine
and near-infrared flux from a surrounding dust torus (Minezaki et al. 2004).
The last point is of particular importance to place useful limits on the mass
of substructure in a foreground gravitational lens.

The paper is organized as follows. In \S 2, we show the target selection and
the observations. In \S 3, the procedure of data reduction is presented to
obtain the mid-infrared fluxes and their ratios among the lensed images.
In \S 4, the implications for the current observational results are discussed
and the conclusions are drawn in \S 5. In what follows, we adopt the set of
cosmological parameters of $\Omega=0.3$, $\Lambda=0.7$, and $h=0.7$ for
all relevant estimations.

\begin{figure*}
\centerline{\includegraphics[width=9cm]{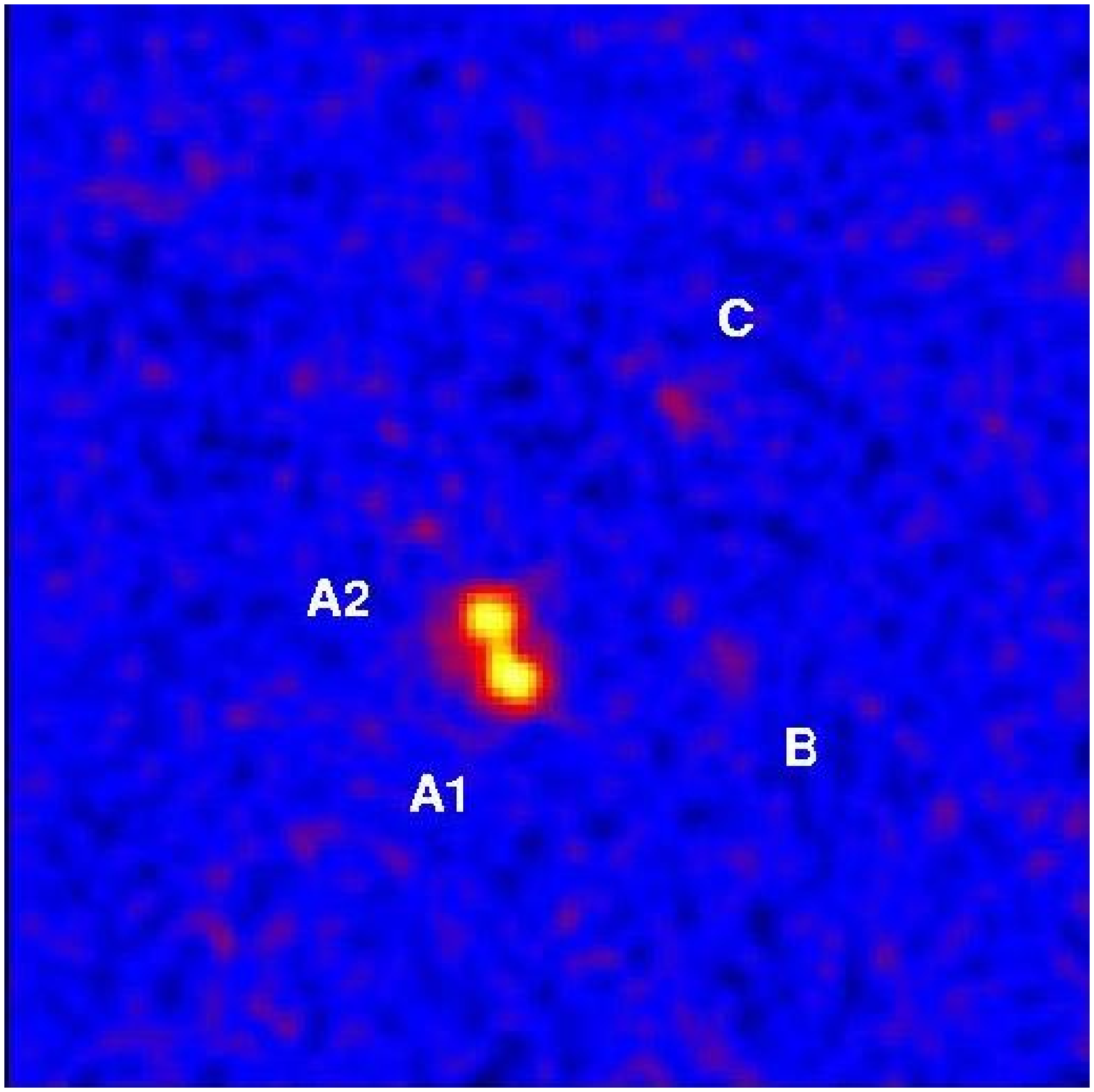}
            \includegraphics[width=9cm]{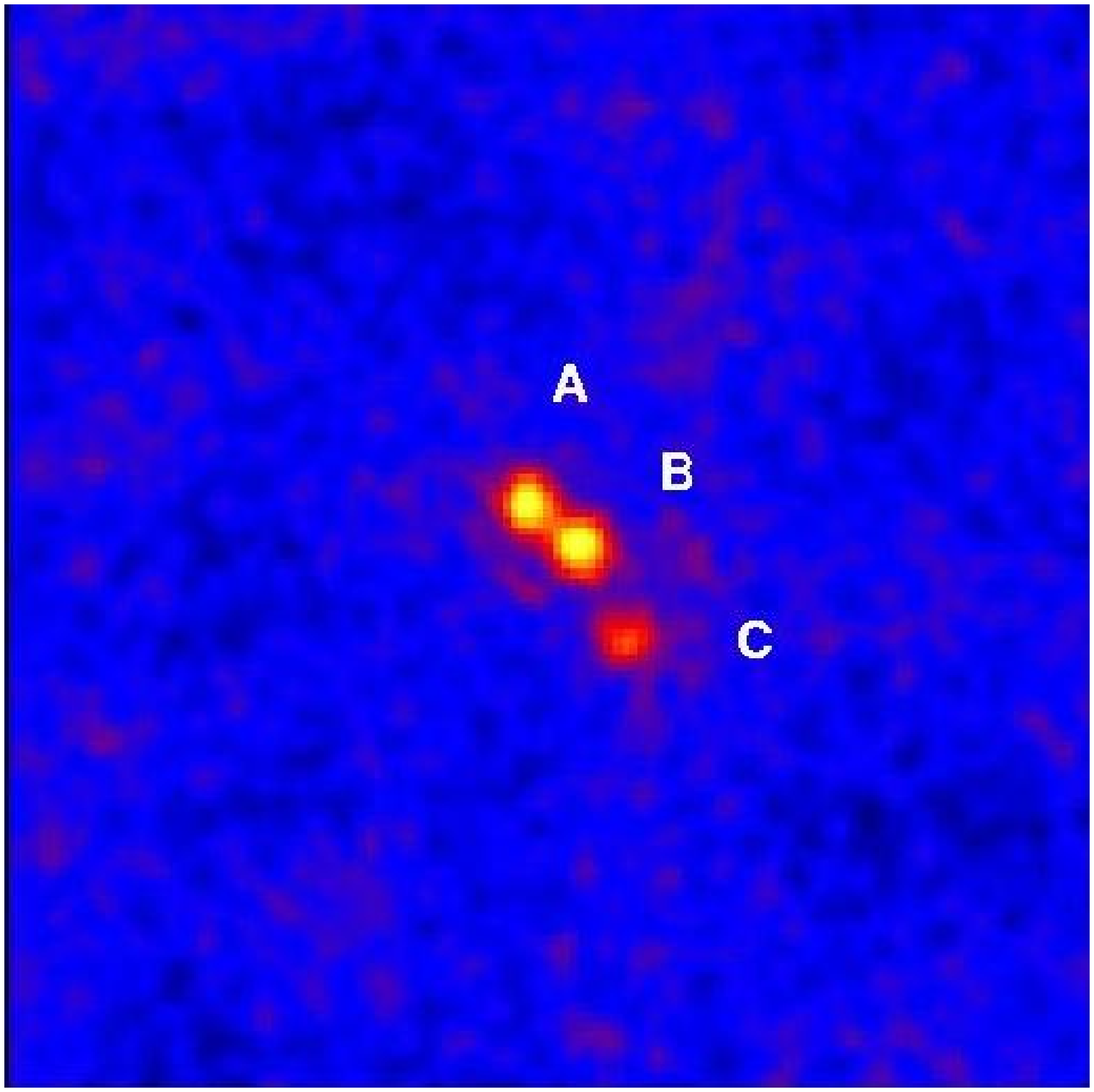}}
\caption{
The quadruple lens systems, PG1115$+$080 (left) and B1422$+$231 (right),
at $11.7\ \mu$m taken with COMICS/Subaru on 2004 May 5 and 6 (UT).
The direction of the images is that the north is up and the east is left,
and the pixel scale of them is $0.\arcsec 065$ pixel${}^{-1}$.
These images have been smoothed with
a Gaussian kernel of $\sigma = 0.\arcsec 065$
in order to improve their visual impression.
}
\end{figure*}

\section{OBSERVATION}

\subsection{Targets}

We select two lens systems with four images, PG1115$+$080 at $z_S=1.72$ and
B1422$+$231 at $z_S=3.62$, renowned for their anomalous image flux ratios.
The lensing galaxies are possibly elliptical galaxies at $z_L = 0.31$ and 0.34,
respectively (Kundi\'c et al. 1997a; Kundi\'c et al. 1997b; Tonry 1998).
The former lens system holds the closely separated pair of images A1
and A2 with a separation of $0.\arcsec 48$,
and this configuration emerges if the QSO is close to and inside
{\it a fold caustic} provided by the foreground lens. The latter shows
the colinear, three highly magnified images, A, B, and C, and this configuration
emerges if the QSO is close to and inside {\it a cusp caustic}. In such lens
systems associated with a fold or cusp caustic, there exists a
universal relation between the image fluxes, i.e., A2$/$A1$=1$
or (A$+$C)$/$B$=1$,
whereas the observed flux ratios violate these rules significantly,
A2$/$A1$=0.64 \pm 0.02$ (Impey et al. 1998) and (A$+$C)$/$B$ = 1.50 \pm 0.01$
(Patnaik et al. 1992). It is worth noting that the flux ratio of A2$/$A1 
in PG1115$+$080 showed little variation with wavelength
from the multiple wavelength observations by Impey et al. (1998),
and that the flux ratios observed in B1422$+$231 by Patnaik et al. (1992)
are based on radio fluxes, so the observed flux ratios of
the concerned images are unlikely to be affected by dust extinction.
Thus, these two lenses are the most representative targets for investigating
the effects of substructure in lensing galaxies.

\subsection{Observation}
The mid-infrared imaging of these lensed QSOs was carried out on the nights of
UT 2004 May 5 and 6, using the cooled mid-infrared camera and spectrometer
(COMICS; Kataza et al. 2000) attached on the Cassegrain focus of
the Subaru telescope.
The field of view is $42\arcsec \times 32\arcsec $ and
the pixel scale is $0.\arcsec 129$ pixel${}^{-1}$.
We used the N11.7 filter, whose effective wavelength and bandwidth
are $\lambda_c = 11.67\ \mu$m and $\Delta \lambda=1.05\ \mu$m, respectively.
The chopping frequency was set $0.45$ Hz with a width of $10\arcsec$, then
the target images were within the field of view for both chopped positions.
In addition to the chop, the telescope position was nodded
several times during the observation.
It was photometric except for the first quarter of both nights,
and the diffraction core of point-spread function (PSF) could be seen.
The FWHM of PSF was $0.\arcsec 33$ at small airmass.
Since we found some degradation of PSF
for the images at large airmass,
we limited the data to those at small airmass,
$\le 1.6$ for PG1115$+$080 and $\le 1.4$ for B1422$+$231.
The total exposure times of the available data were
1.8 hours for PG1115$+$080 and 3.1 hours for B1422$+$231, respectively.
We also observed HD98118 and HD127093
for the photometric standard stars (Cohen et al. 1999).

The images were reduced using IRAF
\footnote{
 IRAF is distributed by the National Optical Astronomy Observatories,
 which are operated by the Association of Universities for Research
 in Astronomy, Inc., under cooperative agreement with the National
 Science Foundation.
}.
An image data file consisted of 208 successive chopped frames
and its exposure time was 200.5 seconds.
First, the sky background of the frame of one chopped position
was subtracted by the frame of the other chopped position,
and flat-field correction was applied using
a sky flat image that was assembled
by combining the frames of the other chopped position
and correcting global variation across the image.
Then two images with exposure time of 100 seconds
were obtained from an image data file
by combining the reduced frames of each chopped position.
After that, in order to reduce residual sky background
around the targets, we fitted it with a low order function
on the square area of $8.3\arcsec \times 8.3\arcsec $
around the target (masking center $3.1\arcsec \times 3.1\arcsec $)
and subtracted it.
The small systematic drift of image positions
due to the telescope tracking error
was fitted with smooth functions
using the image positions of QSO themselves
which were detected marginally
in each 100 seconds-exposure image,
then those images were registered
according to the fitted drift functions.
In order not to degrade the angular resolution
by the sub-pixel shifts during the image registration,
{\em drizzle} (Fruchter \& Hook 2002) task
was used to perform sub-pixel shifts
and re-sample with smaller size of pixels than the original,
where the task parameters were set as
$p=1.0$ (equivalent to shift-and-add)
and $s=0.5$ (re-sampled with half size of pixels).
Finally, all available data were combined.

The resultant mid-infrared ($\lambda = 11.7\ \mu$m) images
of the lensed QSOs, PG1115$+$080 and B1422$+$231,
are presented in Figure 1.
The images presented in the figure have been smoothed with
a Gaussian kernel of $\sigma = 0.\arcsec 065$
in order to improve their visual impression,
although photometries were carried out based on the images
without any smoothing.
As presented in Figure 1,
the lensed images A1 and A2 of PG1115$+$080
and A, B, and C of B1422$+$231 were clearly detected
and well separated from each other.
The minimum image separations, $0.\arcsec 48$ and $0.\arcsec 50$
for these lens systems, respectively,
are $3.0$ and $3.1$ times larger than $\sigma _r$ of PSF
($\sigma_r=0.\arcsec16$, as will be described later).
The faint lensed images B and C of PG1115$+$080
were also detected.

\section{RESULTS}

\subsection{Flux Ratio}

The flux ratios between the lensed images of these QSOs
were estimated by PSF fitting photometry.
We assumed a circular, Gaussian radial profile for the PSF model,
and the relative positions between the images
were taken from 
the CASTLES web site\footnote{http://cfa-www.harvard.edu/glensdata/},
which are based on imaging observation
using the {\it Hubble Space Telescope} (HST).
The free parameters for this fit were
the fluxes of the lensed images,
the positional shift of the whole images,
and $\sigma_r$ for the Gaussian radial profile of PSF.
We used $\phi=1.\arcsec 3$ around each lensed image
for the fitting aperture,
although the estimated flux ratios were almost
independent of the size of the fitting aperture;
using a Gaussian radial profile for the model PSF,
the diffraction core of the real PSF was fitted effectively.
In the case of PG1115$+$080,
we first fit only the images A1 and A2,
then with the derived parameters being fixed
we fit the images A1, A2, B, and C
(where the fluxes of B and C are free parameters).
In the case of B1422$+$231,
we fit only the images A, B, and C
because the faint image D was not detected.

Since the error of flux ratio was not determined
only by photon statistics and detector noise,
and since the fluctuation of sky background was important,
the error of flux ratio was estimated by the following simulation.
First, eight blank-sky areas of $8.3\arcsec \times 8.3\arcsec $
were selected surrounding the objects,
and the model images of targets were added
to the sky areas of the reduced image with exposure time of 100 seconds.
The simulated model image has a Gaussian radial profile
with the best fit parameters of the total flux and $\sigma_r$.
The flux ratio of the simulated model images was also
the same as that of the best-fit to the observation.
Then, those model images were reduced
in the same way as on the real images, and
the PSF fitting photometries were applied to them
to measure flux ratios.
Finally, the error of the observed flux ratio was derived
from the standard deviation of the flux ratios
of the eight simulated images.

\begin{deluxetable}{lllll}
\label{tab1}
\tablecolumns{5}
\tablecaption{Mid-infrared Flux Ratio and Flux of PG1115$+$080}
\tablehead{
\multicolumn{3}{c}{flux ratio} & \multicolumn{2}{c}{flux} \\
\colhead{A2$/$A1} & \colhead{B$/$A1} & \colhead{C$/$A1} & \colhead{A1$+$A2} & \colhead{All} \\
\colhead{} & \colhead{} & \colhead{} & \colhead{(mJy)} & \colhead{(mJy)}}
\startdata
$0.93\pm 0.06$ & $0.16\pm 0.07$ & $0.21\pm 0.04$ & $14.6\pm 1.2$ & $17.5$\tablenotemark{a}
\enddata
\tablenotetext{a}{The total flux of all four images,
based on the flux of A1$+$A2 and the flux ratios listed here.}
\end{deluxetable}

\begin{deluxetable}{lllc}
\label{tab2}
\tablecolumns{4}
\tablecaption{Mid-infrared Flux Ratio and Flux of B1422$+$231}
\tablehead{
\multicolumn{3}{c}{flux ratio} & \colhead{flux} \\
\colhead{(A$+$C)$/$B} & \colhead{A$/$B} & \colhead{C$/$B} & \colhead{A$+$B$+$C} \\
\colhead{} & \colhead{} & \colhead{} & \colhead{(mJy)} }
\startdata
$1.51\pm 0.06$ & $0.94\pm 0.05$ & $0.57\pm 0.06$ & $19.2\pm 2.9$ 
\enddata
\end{deluxetable}

The resultant mid-infrared flux ratios and their one-sigma errors
are presented in Table 1 and 2.
The best-fit values of $\sigma_r $ to the observation
were $0.\arcsec 16$ for both lensed QSOs.
For comparison, Table 3 and 4 show the prediction of a smooth lens model
(Chiba 2002) and the flux ratios in other wavebands reported for the concerned
images\footnote{Errors listed in these tables are, when available, based on
the reported magnitude errors in lensed images or relative errors
in flux ratios.}. 
It follows that
(1) the mid-infrared flux ratio, $r_{IR}$, obtained for A2$/$A1 of PG1115$+$080
is consistent with the prediction of a smooth lens model,
$r_{smooth}$ ($\simeq 1$), whereas the optical flux ratio, $r_{opt}$
($\simeq 0.65$), is systematically smaller than $r_{smooth}$.
(2) the mid-infrared flux ratios for \rm{(A, B, C)} images of B1422$+$231
are virtually consistent with the reported radio flux ratios
(Patnaik et al. 1992), i.e., the mid-infrared flux ratios remain anomalous
as well compared with the prediction of a smooth lens model. These properties
in the flux ratios of both lens systems provide important implications for
the nature of lens substructure as discussed later.

\begin{deluxetable}{lllll}
\label{tab3}
\tablecolumns{5}
\tablecaption{Comparison of Flux Ratios for PG1115$+$080}
\tablehead{\colhead{Wavelength} & 
\colhead{Date} & 
\colhead{A2$/$A1} & \colhead{B$/$A1} & \colhead{C$/$A1}}
\startdata
Lens Model\tablenotemark{a}  &               & 0.92          & 0.22          
    & 0.28             \\
\tableline
11.7 $\mu$m\tablenotemark{b} & 2004.05.05-06 & $0.93\pm0.06$ & $0.16\pm0.07$ 
    & $0.21\pm0.04$    \\
F160W\tablenotemark{c}     & 1997.11.17      & $0.63\pm0.02$ & $0.16\pm0.01$ 
    & $0.25\pm0.01$    \\
F814W\tablenotemark{c}     & 1997.05.17      & $0.67\pm0.02$ & $0.17\pm0.01$ 
    & $0.25\pm0.01$    \\
F555W\tablenotemark{c}     & 1999.03.31      & $0.52\pm0.07$ & $0.15\pm0.02$ 
    & $0.25\pm0.04$    \\
F160W\tablenotemark{d}     & 1997.11.17      & $0.64\pm0.02$ & $0.17\pm0.01$ 

    & $0.26\pm0.01$    \\
F160W\tablenotemark{e}     & 1997.11.17      & $0.67\pm0.04$ & $0.22\pm0.02$ 
    & $0.29\pm0.01$    \\
K'\tablenotemark{f}        & 1999.01.11-13   & 0.67          & 0.19          
    & 0.28             \\
J\tablenotemark{f}         & 1999.01.11-13   & 0.59          & 0.16          
    & 0.24             \\
F785LP\tablenotemark{g}    & 1993.02.18      & 0.71          & 0.18          
    & 0.27             \\
F555W\tablenotemark{g}     & 1993.02.18      & 0.63          & 0.17          
    & 0.23             \\
F785LP\tablenotemark{h}    & 1991.03.03      & $0.70\pm0.01$ & $0.16\pm0.003$
    & $0.26\pm0.01$    \\
F555W\tablenotemark{h}     & 1991.03.03      & $0.66\pm0.03$ & $0.16\pm0.01$ 
    & $0.26\pm0.01$
\enddata
\tablenotetext{a}{Prediction of the smooth lens model by Chiba (2002).}
\tablenotetext{b}{This work.}
\tablenotetext{c}{Taken from the CASTLES web site. The observing date
is taken from the HST data archive.}
\tablenotetext{d}{Impey et al. (1998).}
\tablenotetext{e}{Treu \& Koopmans (2002).}
\tablenotetext{f}{Iwamuro et al. (2000).}
\tablenotetext{g}{Unpublished WFPC1 images analyzed by Impey et al. (1998).}
\tablenotetext{h}{Kristian et al. (1993).}
\end{deluxetable}

\begin{deluxetable}{lllll}
\label{tab4}
\tablecolumns{5}
\tablecaption{Comparison of Flux Ratios for B1422$+$231}
\tablehead{\colhead{Wavelength} & 
\colhead{Date} & 
\colhead{(A$+$C)$/$B} & \colhead{A$/$B} & \colhead{C$/$B}}
\startdata
Lens Model\tablenotemark{a}  &               & 1.25          & 0.76          
    & 0.50             \\
\tableline
11.7 $\mu$m\tablenotemark{b} & 2004.05.05-06 & $1.51\pm0.06$ & $0.94\pm0.05$ 
    & $0.57\pm0.06$    \\
F160W\tablenotemark{c}     & 1998.02.27      & $1.43\pm0.05$ & $0.90\pm0.03$ 
    & $0.53\pm0.02$    \\
F791W\tablenotemark{c}     & 1999.02.06      & $1.57\pm0.13$ & $0.97\pm0.10$ 
    & $0.60\pm0.04$    \\
F555W\tablenotemark{c}     & 1999.02.06      & $1.57\pm0.18$ & $1.02\pm0.14$ 
    & $0.55\pm0.06$    \\
F480LP\tablenotemark{d}    & 1995.04.20      & $1.34\pm0.04$ & $0.79\pm0.02$ 
    & $0.54\pm0.02$    \\
F342W\tablenotemark{d}     & 1995.04.21      & $1.21\pm0.04$ & $0.68\pm0.02$ 
    & $0.53\pm0.02$    \\
Gunn r\tablenotemark{e}    & 1993.04.26      & $1.23\pm0.02$ & $0.75\pm0.015$
    & $0.48\pm0.010$   \\
Gunn g\tablenotemark{e}    & 1993.04.26      & $1.25\pm0.02$ & $0.77\pm0.020$
    & $0.48\pm0.010$   \\
8.4 GHz\tablenotemark{f}   & 1997.06.11-12   & $1.42\pm0.02$ & $0.93\pm0.02$ 
    & $0.49\pm0.01$    \\
5 GHz\tablenotemark{g}     & 1991.08.31      & $1.50\pm0.01$ & $0.98\pm0.01$ 
    & $0.52\pm0.01$    \\
8.4 GHz\tablenotemark{g}   & 1991.06.16      & $1.48\pm0.01$ & $0.97\pm0.01$ 
    & $0.52\pm0.01$
\enddata
\tablenotetext{a}{Prediction of the smooth lens model by Chiba (2002).}
\tablenotetext{b}{This work.}
\tablenotetext{c}{Taken from the CASTLES web site. The observing date is
taken from the HST data archive.}
\tablenotetext{d}{Impey et al. (1996).}
\tablenotetext{e}{Yee \& Ellingson (1994).}
\tablenotetext{f}{Patnaik et al. (1999).}
\tablenotetext{g}{Patnaik et al. (1992).}
\end{deluxetable}

\subsection{Flux and Spectral Energy Distribution}

The fluxes of these lensed QSOs
were estimated by aperture photometry.
We used $\phi=1.\arcsec 3$ for the photometric aperture.
Since the lensed images were so close each other
that their flux distributions were overlapped each other,
we measured the fluxes of
$f$(A1$+$A2) of PG1115$+$080 and $f$(A$+$B$+$C) of B1422$+$231
by connecting the photometric apertures.
The B and C images of PG1115$+$080 were so faint
that aperture photometries for them were not executed.
Instead, the total flux of all images of PG1115$+$080
was estimated from $f$(A1$+$A2)
and the flux ratios we obtained.
After measuring the fluxes within the apertures,
the aperture corrections were applied to estimate the total fluxes.
The amount of corrections was estimated as about 10\%.
The error of flux was estimated
in the same way as on the flux ratio.

The fluxes of the lensed QSOs were calibrated
by comparing the fluxes of standard stars, HD98118 and HD127093.
Since HD127093 is a variable star of irregular type,
it was used only for the estimation of an airmass gradient.
The flux of HD98118 was calculated
by integrating its spectral energy distribution (SED)
presented by Cohen et al. (1999)
with the top-hat function of $\lambda_c=11.67\ \mu$m
and $\Delta \lambda=1.05 \ \mu$m.

The resultant fluxes and their one-sigma errors are also listed
in Table 1 and 2, and our results of mid-infrared fluxes of PG1115$+$080 and
B1422$+$231 are graphically presented in Figure 2, where fluxes in other
wavebands are also shown for comparison
(CASTLES; Barvainis \& Ivison 2002; Iwamuro et al. 2000;
 Impey et al. 1998;  Courbin et al. 1997;
 Christian, Grabtree \& Waddell 1987;
 Impey et al. 1996; Remy et al. 1993;
 Yee \& Bechtold 1996; Yee \& Ellingson 1994;
 Oyabu et al. 2001; Patnaik \& Narasimha 2001;
 Patnaik et al. 1992).
It follows that our mid-infrared fluxes of both QSOs present clear
infrared bumps in their SEDs, thereby indicating that the observed
mid-infrared fluxes of PG1115$+$080 and B1422$+$231 are dominated by
thermal emission of a dust torus surrounding the central black hole
and accretion disk. The SEDs in Figure 2 are compared with
the mean energy distribution (MED) of low-redshift QSOs
by Elvis et al. (1994),
and our mid-infrared fluxes are consistent with the MED,
if the dispersion of SED around the MED is taken into account.
Oyabu et al. (2001) measured the total flux of B1422$+$231
in mid-infrared bands using ISO. The observation date was UT 1997 January 13,
and the reported fluxes were
$5.8$ mJy in LW2($\lambda=6.7\ \mu$m)
and $15.1$ mJy in LW3($\lambda=14.3\ \mu$m).
These fluxes seem to be somewhat fainter than our result,
however these observations are not inconsistent
if the error of our total flux
and the absolute photometric calibration error
of ISO observation (15\% in Oyabu et al. 2001)
are considered.
Therefore, mid-infrared flux variation
was not found between two observations of B1422$+$231.

\begin{figure}
\includegraphics[width=8.5cm,angle=0]{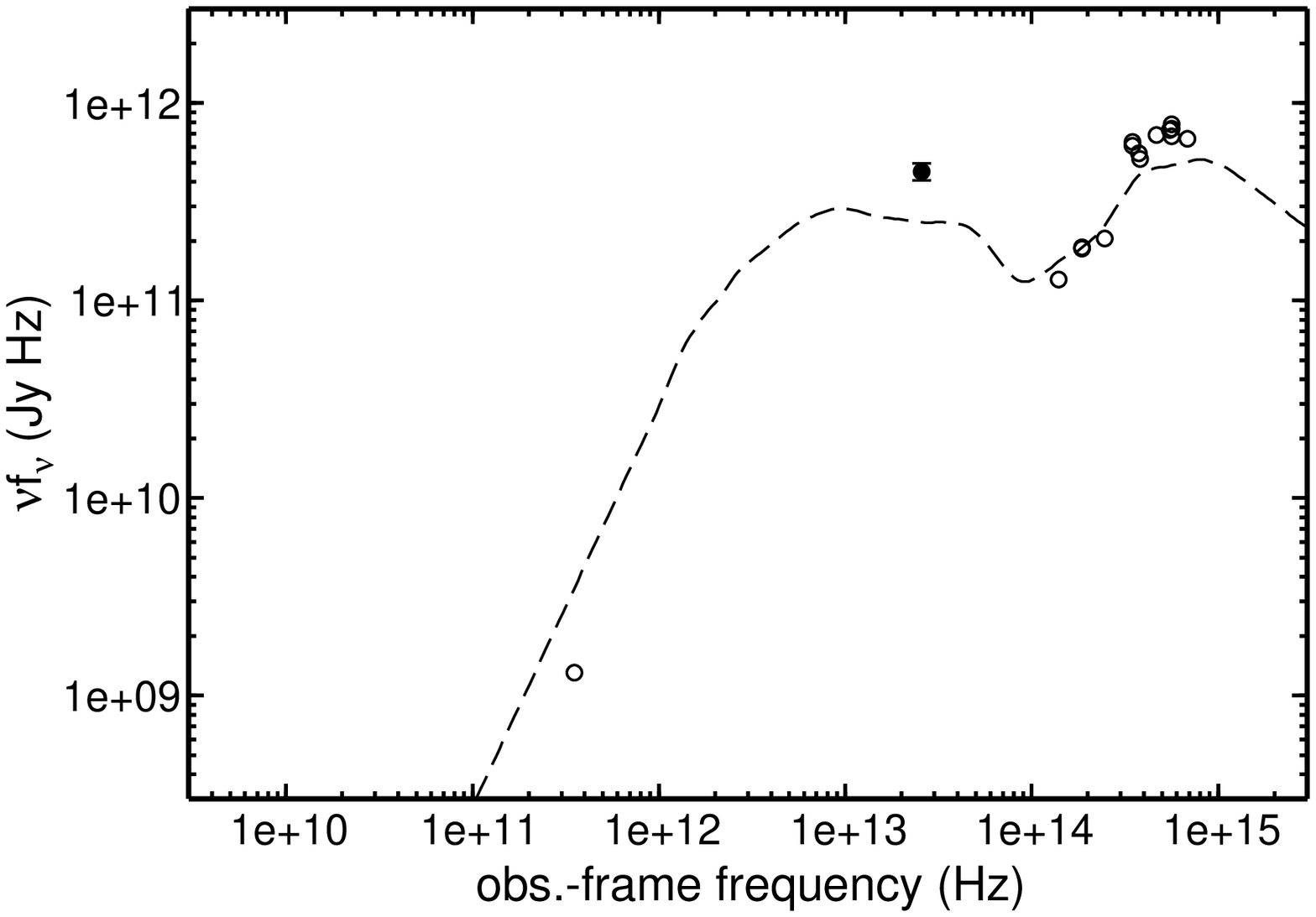}
\includegraphics[width=8.5cm,angle=0]{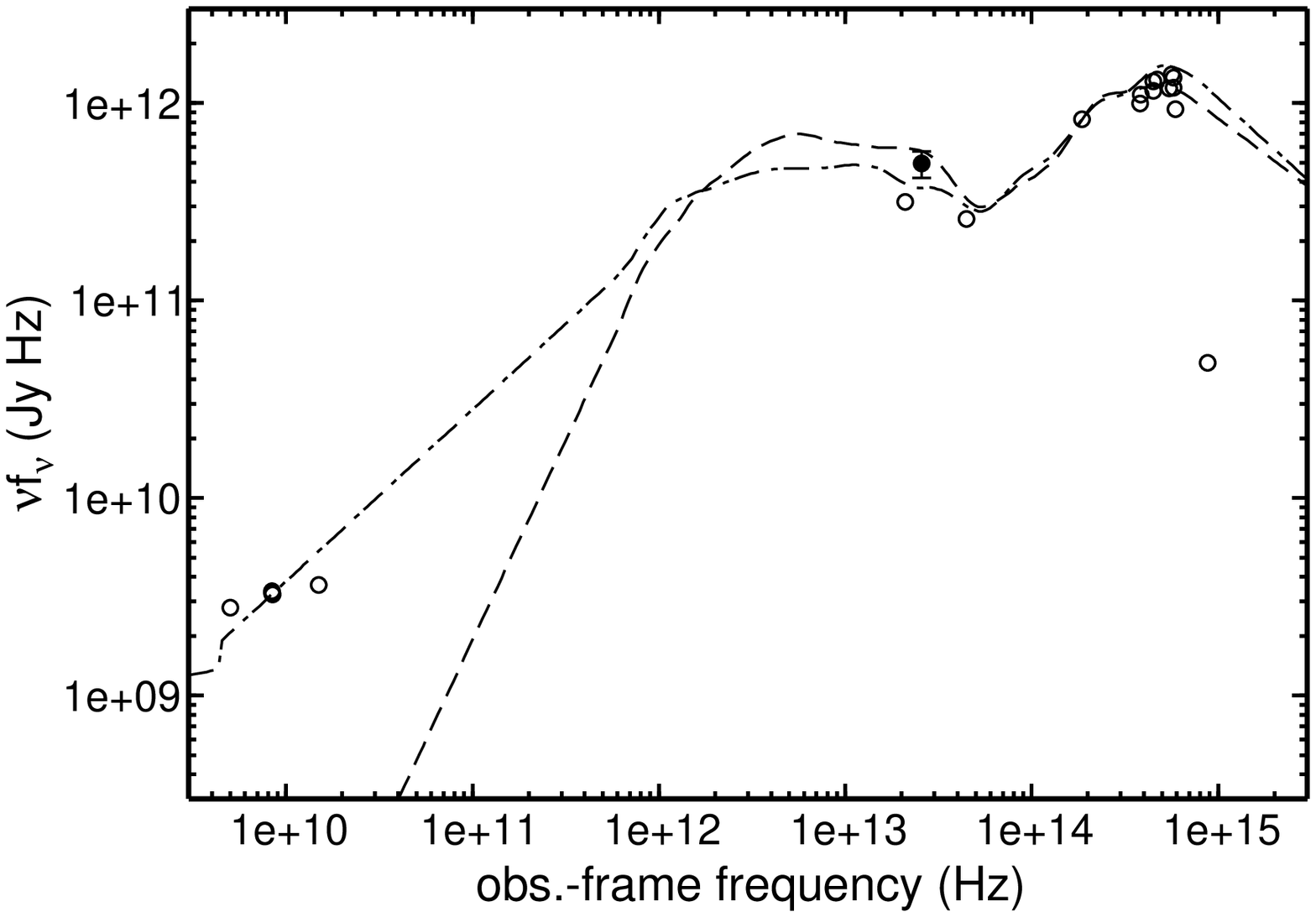}
\caption{The spectral energy distributions of PG1115$+$080 (left) and
B1422$+$231 (right). The open circles are taken from literature, and
the filledcircles are the 11.7 $\mu $m fluxes of this work. The dashed
line is the mean energy distribution (MED; Elvis et al. 1994) of radio
quiet QSOs, and the dash-dotted line is the MED of radio loud QSOs,
both of which are normalized by the F160W fluxes taken from the CASTLES.}
\end{figure}

\section{DISCUSSION}

\subsection{Limits on Substructure Lensing}

The mid-infrared flux ratios presented here set important limits
on the nature of lens substructure,
because the observed flux is dominated by thermal emission of hot dust
at the innermost region of a dust torus,
which is extended and whose size can be estimated based on
the results of the reverberation observations.
Using the lag time $\Delta t$ from the UV-optical continuum flux variation
to the near-infrared flux variation,
which is regarded as the light-travel time
from the central accretion disk to the innermost dust torus,
the radius of a near-infrared emission region is estimated
as $R = c \Delta t$ where $c$ is the speed of light.
More luminous active galactic nuclei (AGN) have larger dust tori,
and the lag time of near-infrared ($K$-band, $\lambda=2.2\ \mu$m) flux variation
can be estimated as $\log \Delta t ({\rm days}) = -2.15 - M_V / 5.0$
where $M_V$ is the absolute magnitude of AGN in $V$ band,
according to the results of dust reverberation observations
compiled by Minezaki et al. (2004).

Based on the total apparent magnitudes of the quadruple images in F160W
taken from the CASTLES web site
(14.9 mag for PG1115$+$080 and 13.3 mag for B1422$+$231)
and the magnification factors due to lensing galaxies
(32.73 and 22.79, respectively, estimated from the lens model of Chiba 2002),
we obtain $M_V = -24.6$ mag and $-28.3$ mag, respectively.
Then, the radii of the rest-frame $K$-band emission regions
are estimated as $R=0.5$ pc and $2.7$ pc
for PG1115$+$080 and B1422$+$231, respectively.
The rest-frame wavelength of the observing $11.7\ \mu$m of PG1115$+$080
is $4.3\ \mu$m, which is somewhat longer than that of $K$ band.
Concerning the contribution to the observed mid-infrared flux of cooler dust,
we estimate the source radius as $R_S\sim 1$ pc.
The source radius of observed mid-infrared flux of B1422$+$231
is estimated as $R_S\simeq 2.7$ pc,
because its observing wavelength of $11.7\ \mu$m
corresponds to the rest-frame wavelength of $2.5\ \mu$m,
almost the same as that of $K$ band.
At last, we convert the linear dimension $R_S$ of a source image
into an angular size of $\theta_S$,
and we obtain $\theta_S \simeq 1 \times 10^{-4}$ arcsec and
$3.7 \times 10^{-4}$ arcsec for PG1115$+$080 and B1422$+$231, respectively.

Next, the characteristic sizes of UV-optical continuum emission regions are
estimated. They would be dependent on QSO luminosity,
and also dependent on observing wavelength
as larger source size is expected at longer wavelength.
We estimate the source sizes for the F160W observations of
PG1115$+$080 and B1422$+$231
as $R_S\sim 6 \times 10^{-3}$ pc and $2\times 10^{-2}$ pc,
yielding $\theta_S \la 7 \times 10^{-7}$ arcsec and $3 \times 10^{-6}$ arcsec,
respectively (see Appendix for details).

Either a star with sub-solar mass (microlensing) or a CDM subhalo with
$M = 10^{7-9} M_\odot$ (millilensing) is responsible for the reported anomalous
flux ratios. To distinguish between these substructure lensing effects, it
is worth estimating an Einstein angle $\theta_E$ to compare with $\theta_S$.
Denoting $M_E$ as a substructure mass inside an Einstein angle, the latter is
expressed as,
\begin{equation}
\theta_E = 8 \times 10^{-7} 
              \left( \frac{M_E}{0.1 M_\odot} \right)^{1/2}
              \ \rm{arcsec} \ ,
\end{equation}
for both lens systems. This equation applies to both a point-mass lens and
spatially extended one, provided a lens is circularly symmetric in projection.
As a reference for the total mass of a spatially extended substructure,
being applicable to a CDM subhalo, we assume a singular isothermal sphere (SIS),
which is parameterized by a one-dimensional velocity dispersion,
$\sigma_\parallel$, or circular velocity, $V_c = \sqrt{2} \sigma_\parallel$,
so that the Einstein angle is
written as $\theta_E \simeq 1 \times 10^{-5} ( V_c / {\rm km~s}^{-1} )^2$
arcsec for both lens systems. Then, the total mass of an SIS substructure inside
radius of 100~pc is given as
$M_{100}^{\rm SIS} = 2.3 \times 10^4 ( V_c / {\rm km~s}^{-1} )^2 \ M_\odot$.

Compared with $\theta_S$ for a surrounding dust torus or continuum-emitting
region, one can immediately draw the conclusion that a star with sub-solar mass
is unable to magnify mid-infrared flux arisen from a dust torus. This is also the
case for low-mass CDM subhalos, provided $\theta_E \ll \theta_S$.
Then, for PG1115$+$080, the reported anomalous flux ratios in the optical band
may be caused by either microlensing of a star with sub-solar mass
or by a very low-mass CDM subhalo with $M_E \la 1.6 \times 10^3 M_\odot$
from the $\theta_E \la \theta_S$ ansatz,
where both of these substructure lenses do not magnify mid-infrared flux
from a dust torus. Tighter mass limits are available, based on the detailed
simulations by Wyithe, Agol, \& Fluke (2002), which showed that
$\theta_E$ must be at least an order of magnitude smaller than $\theta_S$
for no magnification. We then obtain $M_E \la 16 M_\odot$ inside an Einstein
radius and $M_{100}^{\rm SIS} \la 2.2 \times 10^4 M_\odot$ inside
radius of 100~pc if a substructure is modeled as an SIS.
For B1422$+$231, both of the radio and mid-infrared flux ratios appear to be
modified by substructure lensing compared with the prediction of
a smooth lens model; the difference in the most problematic flux ratio, A$/$B,
is about 20~\%. Taking into account $\theta_S$ of a dust torus,
the simple ansatz $\theta_E \ga \theta_S$ for the amplification of
a torus suggests a substructure with $M_E \ga 2.1 \times 10^4 M_\odot$.
Recently, Inoue \& Chiba (2004) investigated the detailed properties of
extended source effects in substructure lensing and showed that even if
a source size is about ten times larger than an Einstein radius of
an SIS-form substructure, the flux ratio is changed by about 20~\% from a
smooth-lens prediction, provided a substructure is just centered at a source
image. Thus, we obtain $M_E \ga 209 M_\odot$ inside an Einstein
radius, whereby a CDM subhalo is more likely, and
$M_{100}^{\rm SIS} \ga 7.4 \times 10^4 M_\odot$ inside radius of 100~pc
if a substructure is modeled as an SIS.

\subsection{Effects of Other Possible Mid-Infrared Sources}

Although the mid-infrared fluxes of both target QSOs appear to be dominated
by thermal emission of a surrounding dust torus, minor fraction of them
might be contributed by the flux of more compact regions than a dust torus,
such as an extension to infrared wavebands of optical continuum emission
from a central accretion disk (Collin \& Hur\'e 2001) or non-thermal emission
from a compact region relating to radio activity
(Neugebauer \& Matthews 1999; Enya et al. 2002). Also,
more spatially extended sources than a dust torus, such as a region of
nuclear starburst and/or QSO's host galaxy as a whole,
are to be taken into account.
We discuss here the effects of possible contribution to mid-infrared fluxes
by both more compact and extended regions than a dust torus.

\subsubsection{Effects of More Compact Sources}

The contributions from the accretion disk
to the mid-infrared flux are estimated
as $\sim 0.06$ for PG1115$+$080
and $\sim 0.2$ for B1422$+$231,
by extrapolating the optical power-law continuum
from the CASTLES F160W fluxes with index
$\alpha _{\nu}=0$ where $f_{\nu}\propto \nu^{\alpha_{\nu}}$
[cf. $\alpha _{\nu}=1/3$ for the standard accretion disk
(Shakura \& Sunyaev 1973)
and $\alpha _{\nu}=-0.44$ from the composite spectrum
of SDSS QSOs (Vanden Berk et al. 2001)].
The contribution estimated by extrapolating
the radio emission for B1422$+$231 is negligibly small
as followed from Figure 2, and PG1115$+$080 is radio-quiet.

For PG1115$+$080, the effects of mid-infrared flux
from a compact region are also invalidated from the following argument.
As a working hypothesis, we suppose
that intrinsic mid-infrared flux prior to lens magnification,
denoted as $f_{IR}^0$,
arises both from a surrounding hot dust torus, $s f_{IR}^0$,
and from a compact region such as an accretion disk,
$(1-s) f_{IR}^0$, where $s$ $(0 \le s \le 1)$
yields the former fraction of the dust torus
in intrinsic mid-infrared flux. Taking a pair of
(A1, A2) images for the flux ratio, suppose that a star
or a low-mass CDM subhalo
causes additional magnification for a compact region with a factor
of $\delta_i^{CR}$ ($i = A1$ for A1 image or $i = A2$ for A2 image),
such that a total magnification factor is given as
$( 1 + \delta_i^{CR} ) \mu_{0,i}$, where $\mu_{0,i}$ denotes
the magnification without a lens perturber. It is also supposed that
mid-infrared flux from a dust torus remains unchanged.
Then, the mid-infrared flux ratio, $r_{IR}$, is given as,
\begin{equation}
r_{IR} = r_{smooth} \times
         \frac{s + (1+\delta_{A2}^{CR}) (1-s)}{
               s + (1+\delta_{A1}^{CR}) (1-s)} \ .
\end{equation}
If mid-infrared emission is supplied
only from the compact region such as an accretion disk ($s=0$),
we obtain $r_{IR} = r_{opt}
\equiv r_{smooth} (1+\delta_{A2}^{CR}) / (1+\delta_{A1}^{CR})$,
which is inconsistent with the current observational result of
$r_{IR} \simeq r_{smooth} > r_{opt}$. Similarly, a more massive substructure
with $M$ being large enough to affect mid-infrared flux from a dust torus
yields $r_{IR}=r_{opt}$ as well and is thus unlikely. Instead, if mid-infrared
emission originates from a dust torus alone ($s=1$), we reproduce the
observational result of $r_{IR} \simeq r_{smooth}$.
Thus, thermal emission of the dust torus
is a main contributor to the observed
mid-infrared flux in PG1115$+$080,
which is consistent with the infrared-bump feature of its SED,
and the lens perturber that produces
the anomalous flux ratios in the optical bands of PG1115$+$080
is either a star with sub-solar mass
or a low-mass CDM subhalo.

For B1422$+$231, contribution of a compact region to the observed mid-infrared
flux is estimated as about 0.2, and this implies, as shown below,
that a lens perturber
ought to be more massive than a star with sub-solar mass.
For instance, we consider the flux ratio A$/$B,
whose value in radio or mid-infrared ranging $0.9 - 1.0$ differs largely
from the smooth lens prediction of 0.76, and this anomaly mainly causes
the deviation from the cusp-caustic relation, (A$+$C)$/$B$=1$.
Then, if we set $s \sim 0.8$ and if a star affects either A or B image
(i.e., $\delta_B^{CR}=0$ or $\delta_A^{CR}=0$),
we obtain $r_{IR} \la 0.8$ from equation (3) for either case as $r_{opt}$
is smaller than 1 (Table 4). This is inconsistent with $r_{IR} \simeq 0.94$
at 3~$\sigma$ level. Also, in this microlensing hypothesis, the observed $r_{IR}$
can be reproduced only when $s = 0.4 - 0.5$ ($50 - 60$~\% contribution from
a compact region to the mid-infrared flux), which is very unlikely.
For $s=1$, microlensing hypothesis is ruled out as well unless otherwise
we expect $r_{IR} = r_{smooth}$.
Curiously, observations in short wavebands [F480LP and F342W (Impey et al. 1996)
and also Gunn $r$ and $g$ (Yee \& Ellingson 1994)] yield the similarity of A$/$B
to the smooth lens prediction, whereas the CASTLES data in
near-infrared band appear to be consistent with the radio and mid-infrared
results (Table 4). This might be caused by differential dust extinction
and/or spatial variation of QSO absorption medium in the foreground lens
but the firm conclusion is yet unavailable.

\subsubsection{Effects of Nuclear Starburst and Host Galaxy}

If the mid-infrared fluxes originate from extended regions,
such as a region of starburst and host galaxy itself,
then the observed images would be largely deformed, possibly in an arc-like
form, by foreground lens galaxies; the deformation of images from
a circular shape is at most by a factor of lensing magnification in the
tangential direction. This conjecture yields upper limits on the sizes of
mid-infrared sources. The observed lensed images of both targets were found
to be a circular shape within the range of their FWHM, $0.\arcsec 38$. Then,
taking into account the largest magnification factors (13.5 for image A1 of
PG1115$+$080, 9.9 for image B of B1422$+$231) obtained in the lens model
of Chiba (2002), the source size $R_S$ for the observed mid-infrared flux must
be less than $\sim 240$~pc for PG1115$+$080 and $\sim 280$~pc for B1422$+$231.

Effects of starburst activity associated with QSOs can be assessed
by the excess of far-infrared dust radiation as obtained from millimeter
and/or submillimeter observations (Carilli et al. 2001; Omont et al. 2001).
For PG1115$+$080, such an excess is unidentified (Barvainis \& Ivison 2002)
as shown in Figure 2. Then, we estimate the possible contribution of nuclear
star formation activity to the observed mid-infrared flux using
the polycyclic aromatic hydrocarbon (PAH) emission feature at 3.3~$\mu$m.
It is known as an indicator of star formation activity, and is located
near the rest-frame wavelength of 4.3~$\mu$m for the observed 11.7~$\mu$m.
Based on the reported equivalent widths of the PAH emission
for nuclear regions of Seyfert 1 ($\la 1 - 10$~nm: Imanishi \& Wada 2004)
and starburst galaxies
($\sim 100$~nm: Imanish \& Dudley 2000), it is found that the contribution
of star formation activity to the currently observed flux is only $0.01 - 0.1$
or less, thereby being unimportant for our results.
On the other hand, for B1422$+$231, no millimeter or submillimeter
observations have been conducted, 
so the possibility of nuclear starburst in this system is yet to be settled.
If additional mid-infrared flux from nuclear starburst is included in the
observed flux, then the anomaly of its flux ratio as reported here would be
caused by a more massive substructure than that based on the dust torus size,
because the nuclear starburst region would be extended on a scale of
$\sim 100$~pc or more. In that case,
the possibility of a CDM subhalo to induce the anomalous flux ratio is made
more robust.

Effects of host galaxies appear to be minor, as judged from the smallness of the
size bounds described above. Using the results of the HST imaging observation for
host galaxies of QSOs at $z \sim 0.4$ (Floyd et al. 2004)
and the color information for low-redshift QSOs' hosts
(Jahnke, Kuhlbrodt, \& Wisotzki 2004; Hunt, Giovanardi, \& Helou 2002),
it is found that the contribution of host galaxies to the observed
mid-infrared fluxes are only less than 0.1~\%.


\subsection{Substructure Lensing in PG1115$+$080}

As argued in \S 4.1, the observed mid-infrared flux ratio A2$/$A1$ \simeq 1$
for PG1115$+$080 implies that its anomalous flux ratio in optical and
near-infrared bands (Table 3) may be induced by either a star with sub-solar
mass or a low-mass CDM subhalo with
$M_{100}^{\rm SIS} \la 2.2 \times 10^4 M_\odot$ if modeled as an SIS; our current
observations do not distinguish between both possibilities. A possible method to
settle this issue is to consider the rapid time variability of these images
caused by microlensing if a star is involved in flux anomaly.
The time variability of lensed QSO emissions can be induced by relative motion
between the source and lens (e.g., Schneider, Ehlers, \& Falco 1992).
For the purpose of simplification, we assume that this relative motion is
dominated by transverse velocity ($v_\perp$) of a substructure lens, i.e.,
the sum of its peculiar velocity inside a galaxy and the motion of the galaxy.
Tonry (1998) showed that the lensing galaxy of PG1115$+$080 has the internal
velocity dispersion of $281 \pm 25$ km~s$^{-1}$ and that the galaxy belongs to
a group with a dispersion of $326$ km~s$^{-1}$, so for simplicity we adopt
$v_\perp \sim 600$ km~s$^{-1}$ for the following estimates.

A useful measure for the time variability is the time required for a star to
cross its Einstein radius with velocity $v_\perp$. For PG1115$+$080, we obtain
\begin{eqnarray}
t_E &=& (1+z_L) \frac{D_{OS} \theta_E}{v_\perp} \frac{D_{OL}}{D_{OS}} \nonumber \\
    &\sim& 7.8 \left( \frac{M}{0.1 M_\odot} \right)^{1/2}
           \left( \frac{v_\perp}{600 \rm{km~s}^{-1}} \right)^{-1} \ \rm{yr} \ ,
\end{eqnarray}
where $D_{OL}$ and $D_{OS}$ are the angular diameter distances to the lens and
source, respectively, and the second equation is derived for the point-mass
lens model. Another characteristic timescale is the time required for a caustic
to cross over a source size $R_S$:
\begin{eqnarray}
t_C &=& 2 (1+z_L) \frac{R_S}{v_\perp} \frac{D_{OL}}{D_{OS}} \nonumber \\
    &\sim& 13.8 \left( \frac{R_S}{6 \times 10^{-3}~\rm{pc}} \right)
           \left( \frac{v_\perp}{600 \rm{km~s}^{-1}} \right)^{-1} \ \rm{yr} \ ,
\end{eqnarray}
where we use the estimated source size of an optical continuum emission region
of PG1115$+$080 using the F160W observation. Note that the size of a UV
continuum emission region is expected to be smaller by some factor as argued
in Appendix, so the timescale can be shorter when observed in shorter wavebands.

Although more realistic estimates must consider a complicated caustic network
involved in a star field rather than a single star, the fact that the optical
flux ratio for A1 and A2 images of PG1115$+$080 has remained basically unchanged
over the past decade (Table 3) seems to be consistent with the implications from
the above rough estimates, i.e, the time variability induced by microlensing is
yet difficult to assess because of its rather long duration.
However, somewhat shorter and low-amplitude variation in these images would be
yet possible, especially in shorter observing wavebands, owing to the effects of
a caustic network in the microlensing hypothesis. In contrast,
the time variability is extremely long in the millilensing hypothesis
and even for less massive CDM subhalos only a single event of flux variation
may be detected because of their much smaller number than stars.
Thus, tighter limits for substructure in the lens of PG1115$+$080 may be set
by a long-term monitoring of A1 and A2 images or that of detailed spectral
features of their emission lines
(Narasimha \& Srianand 2000; Moustakas \& Metcalf 2003).

\section{CONCLUSION}

We have presented mid-infrared imaging at 11.7 $\mu$m for the quadruple lens
systems, PG1115$+$080 and B1422$+$231, which are characterized by the
anomalous flux ratios among their lensed images in other wavebands.
The lensed images of these QSOs have
successfully been separated from each other owing to the diffraction-limited
resolution of the COMICS/Subaru images with ${\rm FWHM}=0.\arcsec 38$.
Our calibration of the mid-infrared fluxes and flux ratios among different
lensed images have revealed that (1) the observed mid-infrared fluxes provide
clear infrared bumps in the SEDs of the target QSOs, indicating that the observed
mid-infrared fluxes originate from a hot dust torus around a QSO nucleus,
(2) the mid-infrared flux ratio A2$/$A1 of PG1115$+$080 is virtually consistent
with smooth lens models but inconsistent with the optical flux ratio,
(3) the mid-infrared flux ratios among \rm{(A, B, C)} of B1422$+$231 are
in good agreement with the radio flux ratios.
Based on the possible dimensions of dust tori around
these QSOs estimated from the relation of dust reverberation mapping, these
observational results imply that the origin of anomalous optical
or radio flux ratios for these QSOs may be caused by either of a star with
sub-solar mass or a low-mass CDM subhalo with an SIS model of
$M_{100}^{\rm SIS} \la 2.2 \times 10^4 M_\odot$ inside radius of 100~pc
for PG1115$+$080 and a CDM subhalo with
$M_{100}^{\rm SIS} \ga 7.4 \times 10^4 M_\odot$ for B1422$+$231,
respectively.

Assessing the presence of many dark subhalos in a galaxy is a difficult task
because of their dark nature but is important for understanding their crucial
roles in the formation processes of visible parts of the Universe as well as
for testing the currently standard CDM scenario itself. Lensed QSOs with
anomalous flux ratios have been, so far, possible sole sites for the
investigation of substructure in their foreground lenses. Further mid-infrared
observations of other such lenses, especially radio-quiet QSOs, would provide
useful insight into the nature of substructure lensing.

\acknowledgments
We thank Takuya Fujiyoshi and Shigeyuki Sako for their expert assistance
during our observing runs with COMICS/Subaru. We are also grateful
to Dr. D. Narasimha for his invaluable comments on this work and
discussion. This work has been supported in part by a Grant-in-Aid for
Scientific Research (15540241) of the Ministry of Education, Culture,
Sports, Science and Technology in Japan.


\appendix
\section{The source size of an optical continuum emitting region}
Since the UV and optical continuum emission of AGNs are considered to
originate from the accretion disk around the central black hole,
we estimate the source size by a simple model of the accretion disk.
Assuming a temperature profile as $T\propto R^{-3/4}$
($R$ is radius of an annulus of the disk),
which is expected from the standard accretion disk model
or the irradiation disk model illuminated by a central source,
the disk spectrum is given by
\begin{equation}
f_\nu = 11.2\ {\rm Jy}\ \left( \frac{R}{{\rm light\ days}} \right)^2
\left( \frac{D}{{\rm Mpc}} \right)^{-2}
\left( \frac{\lambda}{\mu {\rm m}} \right)^{-3}
\left( \frac{X}{4} \right)^{-8/3} \cos i \ ,
\end{equation}
which is derived based on the equation (2) of Collier et al. (1999).
$D$ is the distance to the AGN and $X$ is a parameter determined by
a disk model, estimated as $X\la 4$ by Collier et al. (1999).
$i$ is the disk inclination angle;
$\cos i\la 1$ is expected according to the unified scheme
(e.g. Antonucci 1993)
because both target QSOs are type 1 AGNs
and face-on view is expected.
Thus ignoring parameters of $X$ and $i$,
the relation between the wavelength-dependent source radius
and the luminosity is derived as
\begin{equation}
R(\lambda ) = 2.7\times 10^{-2} \ {\rm pc}
\ \left( \frac{\lambda L_{\lambda }(0.51\mu {\rm m})}{10^{46}\
 {\rm ergs\ s^{-1}}} \right)^{1/2}
\left( \frac{\lambda}{\mu {\rm m}} \right)^{4/3}\ ,
\end{equation}
using $f_{\nu }\propto \nu ^{1/3}$ as calculated
from a temperature profile of $T\propto R^{-3/4}$.
Accroding to this equation, the source size was estimated 
from the optical luminosity and the rest-frame wavelength.


\end{document}